\documentclass[journal=jacsat,manuscript=article]{achemso}

\usepackage{chemformula} 
\usepackage[T1]{fontenc} 
\usepackage{graphicx} 
\usepackage{amsmath, amssymb}
\usepackage{amsfonts}
\usepackage{dcolumn} 
\usepackage{bm} 
\usepackage[utf8]{inputenc}
\usepackage{xcolor}
\usepackage{hyperref} 

\author{Jon Lasa-Alonso}
\affiliation{Centro de Física de Materiales (CSIC-UPV/EHU), Paseo Manuel de Lardizabal 5, 20018 Donostia-San Sebastian, Spain}
\alsoaffiliation{Donostia International Physics Center, Paseo Manuel de Lardizabal 4, 20018 Donostia-San Sebastian, Spain}
\email{jonQNanoLab@gmail.com}
\author{Diego Romero Abujetas}
\affiliation{Donostia International Physics Center, Paseo Manuel de Lardizabal 4, 20018 Donostia-San Sebastian, Spain}
\alsoaffiliation{Instituto de Estructura de la Materia (CSIC), Calle Serrano 113 bis- 119-121-123, 28006 Madrid, Spain}
\alsoaffiliation{Department of Materials Science and Engineering, Stanford University, 496 Lomita Mall, Stanford, CA 94305, USA}
\author{Álvaro Nodar}
\affiliation{Centro de Física de Materiales (CSIC-UPV/EHU), Paseo Manuel de Lardizabal 5, 20018 Donostia-San Sebastian, Spain}
\alsoaffiliation{Donostia International Physics Center, Paseo Manuel de Lardizabal 4, 20018 Donostia-San Sebastian, Spain}
\author{Jennifer A. Dionne}
\affiliation{Department of Materials Science and Engineering, Stanford University, 496 Lomita Mall, Stanford, CA 94305, USA}
\author{Juan Jose Sáenz}
\affiliation{Donostia International Physics Center, Paseo Manuel de Lardizabal 4, 20018 Donostia-San Sebastian, Spain}
\alsoaffiliation{IKERBASQUE, Basque Foundation for Science, Maria Diaz de Haro 3, 48013 Bilbao, Spain}
\author{Gabriel Molina-Terriza}
\affiliation{Centro de Física de Materiales (CSIC-UPV/EHU), Paseo Manuel de Lardizabal 5, 20018 Donostia-San Sebastian, Spain}
\alsoaffiliation{Donostia International Physics Center, Paseo Manuel de Lardizabal 4, 20018 Donostia-San Sebastian, Spain}
\alsoaffiliation{IKERBASQUE, Basque Foundation for Science, Maria Diaz de Haro 3, 48013 Bilbao, Spain}
\author{Javier Aizpurua}
\affiliation{Centro de Física de Materiales (CSIC-UPV/EHU), Paseo Manuel de Lardizabal 5, 20018 Donostia-San Sebastian, Spain}
\alsoaffiliation{Donostia International Physics Center, Paseo Manuel de Lardizabal 4, 20018 Donostia-San Sebastian, Spain}
\alsoaffiliation{IKERBASQUE, Basque Foundation for Science, Maria Diaz de Haro 3, 48013 Bilbao, Spain}
\author{Aitzol García-Etxarri}
\affiliation{Donostia International Physics Center, Paseo Manuel de Lardizabal 4, 20018 Donostia-San Sebastian, Spain}
\alsoaffiliation{IKERBASQUE, Basque Foundation for Science, Maria Diaz de Haro 3, 48013 Bilbao, Spain}

\title{Surface-Enhanced Circular Dichroism spectroscopy on periodic dual nanostructures}

\begin{document}

\begin{tocentry}

Some journals require a graphical entry for the Table of Contents.
This should be laid out ``print ready'' so that the sizing of the
text is correct.

Inside the \texttt{tocentry} environment, the font used is Helvetica
8\,pt, as required by \emph{Journal of the American Chemical
Society}.

The surrounding frame is 9\,cm by 3.5\,cm, which is the maximum
permitted for  \emph{Journal of the American Chemical Society}
graphical table of content entries. The box will not resize if the
content is too big: instead it will overflow the edge of the box.

This box and the associated title will always be printed on a
separate page at the end of the document.

\end{tocentry}

\begin{abstract}
   Increasing the sensitivity of chiral spectroscopic techniques such as circular dichroism (CD) spectroscopy is a current aspiration in the research field of nanophotonics. Enhancing CD spectroscopy depends upon of two complementary requirements: the enhancement of the electromagnetic fields perceived by the molecules under study and the conservation of the helicity of those fields, guaranteed by duality symmetry. 
   In this work, we introduce a systematic method to design nanostructured dual periodic photonic systems capable of enhancing molecular CD spectroscopy resonantly. As an illustration, we engineer a dual 1D silicon nanoparticle array and show that its collective optical modes can be efficiently employed to resonantly enhance by two orders of magnitude the local density of optical chirality and, thus, the CD signal obtained from a given molecular sample on its vicinity. 
  \end{abstract}

\section{\label{sec:level1} Introduction}
Circular dichroism (CD) spectroscopy is a widely used experimental technique which enables the characterization of chiral molecular samples. This technique, which measures the differential absorption of molecules to left- and right-handed circularly polarized light (CPL) excitation, permits, for instance, to characterize the secondary structure of biomolecules and to determine the enantiopurity of pharmaceutical drugs ~\cite{Barron}. Despite its wide applicability, the sensitivity of CD measurements is quite limited due to the weak nature of chiral light-matter interactions. Nanostructured materials have been instrumental enhancing the sensitivity of other molecular spectroscopic techniques such as Surface-Enhanced Raman Scattering (SERS) and Surface-Enhanced Infrared Absorption (SEIRA) spectroscopy, increasing their performance several orders of magnitude \cite{SEIRA1, SEIRA2, SERS1, SERS2, SERS3}. Similarly, increasing the sensitivity of CD spectroscopy with the aid of nanostructures would be highly desirable and it is a research objective which has been pursued in the recent past.

Since the seminal contributions of Y. Tang and A. E. Cohen~\cite{tang_cohen}, several nanophotonic platforms have been proposed to augment the sensitivity of CD experiments through the local and global enhancement of $C(\mathbf{r})$, the local density of electromagnetic chirality. To mention a few, high refractive index nanoparticles~\cite{PRB, CDEnh2, OldACSAitzol, Curto, Vladimiro, IvanACS}, plasmonic chiral~\cite{PRXDesignPples, fCD3000} and nonchiral nanoantennas~\cite{CDEnh1, MicOrCD, fCD3000}, optical waveguides~\cite{Vazquez} and metasurfaces~\cite{ACSDionne, NanophotonicPlatforms, Quidant} have been explored as enablers of surface-enhanced circular dichroism and other chiral spectroscopy techniques. 

Globally enhancing CD spectroscopy around a nanostructure necessitates duality symmetry in the design of the device so that the helicity of the scattered fields is conserved ~\cite{PRLMolina}. Unfortunately, common nanophotonic platforms can only guarantee this symmetry at particular, non-resonant, excitation wavelengths. This limits the efficiency of light-matter interactions between the scattered fields and the molecules under study, limiting the magnitude of the achievable enhancements in CD spectroscopy. 

In this work, we present a novel approach to design helicity preserving and optically resonant nanostructured devices to enhance the CD signal exploiting the lattice modes of periodic arrays of dual high refractive index nanoparticles. Starting from the well-known example of a silicon nanosphere capable of enhancing CD spectroscopy by a factor of 10~\cite{PRB}, we first characterize the resonances of an infinite 1D array of such nanoparticles. We observe that these systems support two different types of emerging lattice resonances: far-field diffractive modes and a non-diffractive lattice resonance mediated by mid- and near-field interactions. Secondly, we find that in finite particle arrays finite chain modes can also be observed and exploited. 

We show that all of these helicity preserving resonant modes can globally enhance the sensitivity of CD spectroscopy over two orders of magnitude, providing one additional order of magnitude enhancement over the response of a single, non-resonant nanoparticle ~\cite{PRB}. Even if the maximum enhancement is given for the far-field diffractive modes, we find that enhancements obtained for the rest of the modes are comparable.

\section{\label{sec:level2} Enhanced chiral light-matter interactions and helicity conservation}

For molecules illuminated by a left ($+$) or right ($-$) handed circularly polarized plane wave in vacuum, the CD signal of a molecule, $\text{CD}_{cpl}$, can be expressed as \cite{tang_cohen, PRB} 
\begin{equation}
\text{CD}_{cpl}=-\frac{4}{\varepsilon_0}\text{Im}(G)|C_{cpl}|,
\end{equation}
where $G$ is the molecular chiral polarizability, $C_{cpl}=\pm \frac{\varepsilon_0\omega}{2c}E_0^2$ the local density of chirality of a plane wave in vacuum, $c$ the speed of light, $\omega$ the angular frequency, $E_0$ the amplitude of the incoming electric field and $\varepsilon_0$ the permittivity of free space. Ref.~8 concluded that in the presence of nonchiral antennas, the CD signal could be locally expressed as
\begin{equation}
\text{CD} = -\frac{4}{\varepsilon_0}\text{Im}(G)C(\mathbf{r}),
\label{CD_C}
\end{equation}
with $C(\mathbf{r}) = -\frac{{\omega}}{2c^2}\text{Im}\Big( \mathbf{E}(\mathbf{r})^*\cdot \mathbf{H}(\mathbf{r}) \Big)$  being the local density of optical chirality\cite{tang_cohen}, where $\mathbf{E}(\mathbf{r})$ and $\mathbf{H}(\mathbf{r})$ are the local electric and magnetic fields, respectively. Although $G$ is a fixed molecular parameter, $C(\mathbf{r})$, can be in principle engineered and enhanced in the presence of optical antennas. Importantly, in the presence of nonchiral antennas one can define a local enhancement factor such that $\text{CD}=f_\text{CD}\text{CD}_{cpl}$ where

\begin{equation} \label{fCD}
f_\text{CD}=\frac{\text{CD}}{\text{CD}_{cpl}}=\frac{C(\mathbf{r})}{|C_{cpl}|}=-\frac{Z_0}{E_0^2}\text{Im}\Big(\mathbf{E}(\mathbf{r})^*\cdot \mathbf{H}(\mathbf{r})\Big),
\end{equation}
$Z_0$ being the impedance of vacuum.

Contrary to what could be intuitively expected, the local density of 
chirality, $C(\mathbf{r})$, is not necessarily related to the local handedness of the field. Nevertheless, as shown in Eq. (\ref{CD_C}), the CD signal is directly proportional to $C(\mathbf{r})$ in the presence of nonchiral environments. Therefore, an essential element to enhance the CD signal is to engineer the local fields, so that the overall sign of $C(\mathbf{r})$ (and, thus, also the sign of $f_{\text{CD}}$) is preserved, while its magnitude is maximized. Our work is focused on finding a nanostructure which, when illuminated by CPL of a given handedness, can enhance the absolute value of $C(\mathbf{r})$ while locally preserving its sign in space. In such a scenario, molecules in the vicinity of our optical resonator will interact with the local field in a way that always contributes positively to CD signal. This allows for a strict spatial control on the molecular absorption rates and facilitates the practical measurement of the field-enhanced CD signal.

In the case of monochromatic fields, the electromagnetic density of chirality  is  intimately related to the local density of helicity~\cite{barnett, cameron, nieto2017chiral, Lisa}. 
The helicity operator can be expressed as
\begin{equation}
\Lambda = \frac{\mathbf{J} \cdot \mathbf{P}}{|\mathbf{P}|} = \frac{1}{k}\nabla \times
\label{helicity}
\end{equation}
where $\mathbf{J}$ is the total angular momentum of light, $\mathbf{P}$ is the linear momentum and the second equality in Eq. \eqref{helicity} is only held for monochromatic fields \cite{Messiah}. The helicity operator is the generator of dual transformations, i.e. rotations of the electric and magnetic fields~\cite{PRLMolina}. Hence, the helicity is a conserved magnitude in the interaction of light with systems symmetric under the exchange of electric and magnetic fields, the so-called dual systems.

In single particles with a dominant dipolar response, duality symmetry is fulfilled for wavelengths which meet the first Kerker's condition, in other words, for wavelengths where the electric and magnetic polarizabilities of the nanostructure are equal ($\alpha_e=\alpha_m$) \cite{nieto2011angle,Mole_k,Zambrana, Aitzol_moebius,olmos2019enhanced}. This condition can be achieved for particular wavelengths of light and materials~\cite{strong_magnetic}. Similarly, in a collection of nanoparticles, duality is a symmetry of the whole system if every individual particle satisfies the first Kerker's condition~\cite{PRLMolina,MikolajPRL} 

Helicity conservation in such scenarios allows us to consider the following expressions for the local electric and magnetic fields\cite{MikolajPRL}
\begin{align}
\label{DualE}
\Lambda \mathbf{E}(\mathbf{r}) &=  \frac{1}{k}\nabla \times \mathbf{E}(\mathbf{r}) = p\mathbf{E}(\mathbf{r})\\
\label{DualH}
\Lambda \mathbf{H}(\mathbf{r}) &=  \frac{1}{k}\nabla \times \mathbf{H}(\mathbf{r}) = p\mathbf{H}(\mathbf{r}),
\end{align}
where the helicity eigenvalue, $p = \pm 1$. From Eqs. \eqref{DualE}-\eqref{DualH} and Maxwell's equations one obtains $\mathbf{H}(\mathbf{r}) = \frac{-i p k}{\omega\mu_0}\mathbf{E}(\mathbf{r})$. Applying this expression to Eq. \eqref{fCD}, the local CD enhancement factor for dual systems can be reformulated as:

\begin{equation}
\label{fCDdualE}
f^{dual}_{\text{CD}} = \frac{1}{E_0^2}|\mathbf{E}(\mathbf{r})|^2.
\end{equation}

On the one hand, Eq. \eqref{fCDdualE} confirms that under duality symmetry, $C(\mathbf{r})$ does not change sign locally upon scattering. Most importantly, this result predicts that we can enhance the local CD experimental signal on a dual structure as far as we are able to enhance the fields close to our optical system. Finally, this expression comes into agreement with other results in the literature, which express the chirality in terms of the Riemann-Silberstein vectors~\cite{RS, IvanACS}.

\section{\label{sec:level3}Lattice resonances}
\begin{figure}[ht]
	\centering
	\includegraphics[width=1\textwidth, scale=0.6]{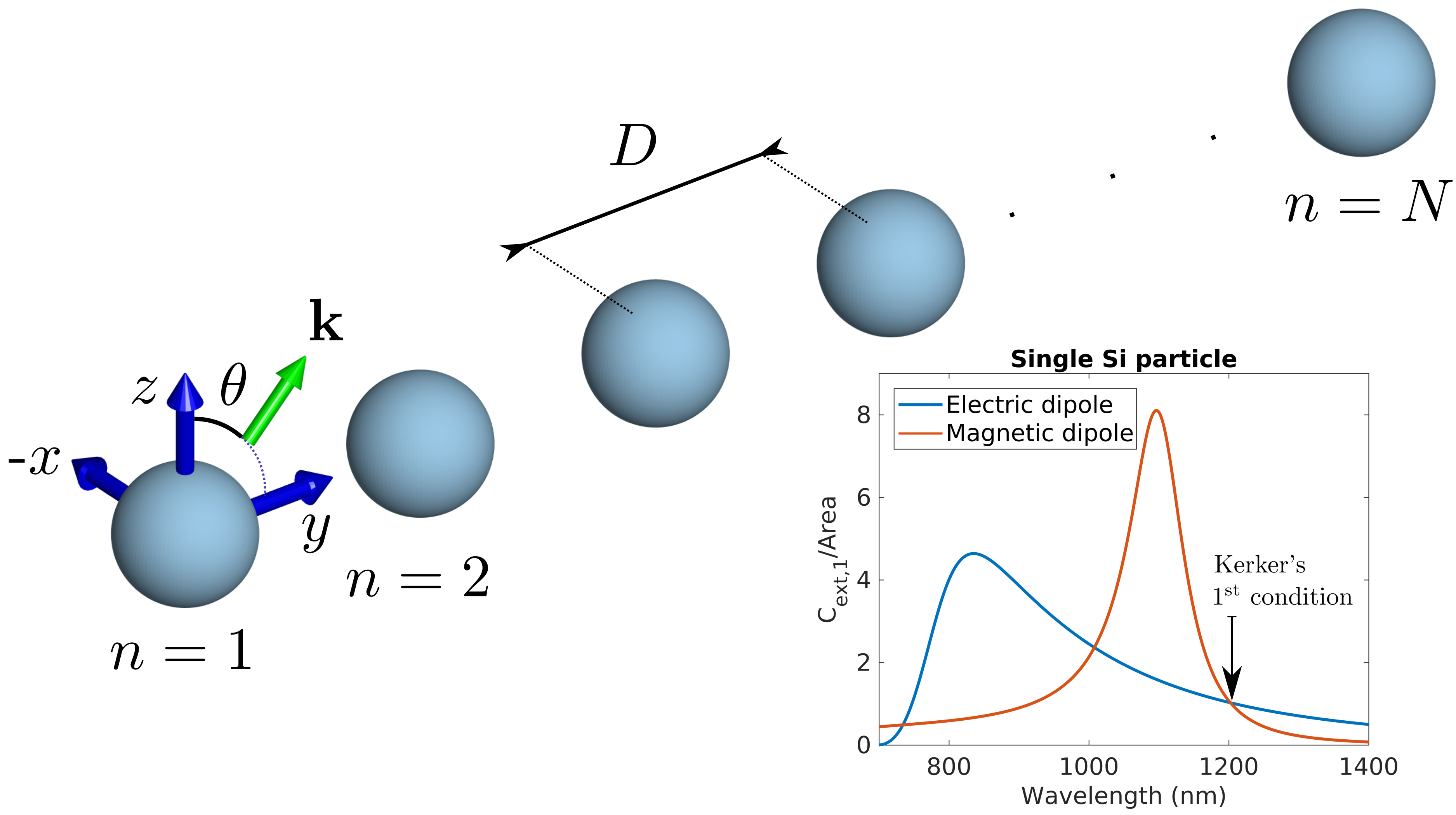}
	\caption{Sketch of the setup under study: a group of nanoparticles arranged in a 1D periodic arrangement. In the bottom right part, the extinction cross section of a single isolated silicon nanoparticle in our system.}
	\label{Array}
\end{figure}

As we determined in the previous section, dual systems allow for the enhancement of $f_\text{CD}$ while keeping its local sign unaltered.  Nevertheless, duality symmetry constriction restricts the chirality enhancement to specific wavelengths which fulfill Kerker's first condition. Typically, this wavelength does not coincide with resonant modes of optical cavities and resonators. As a consequence, the electromagnetic fields used in typical surface-enhanced CD devices are not very intense on isolated nanoparticles. Previous results for single high refractive index nanoparticles \cite{PRB} predict a modest enhancement of one order of magnitude in the molecular CD signal. In order to overcome this limitation we propose to exploit lattice resonances arising in ordered collections of similar dual, non-resonant, individual dielectric nanoparticles. For the sake of clarity, we restrict our study to 1D periodic arrangements of silicon nanoparticles. Nevertheless, ideas presented in this article can be easily generalized for 2D or 3D particle arrays.

Lattice resonances are high-quality factor collective excitations of periodically arranged groups of nanoparticles. Equation \eqref{fCDdualE} indicates that on any dual system $f_\text{CD}$ scales with the square of the field enhancement that is produced. Thus, our hypothesis is that although the individual nanoparticles are non-resonant at the first Kerker's condition, their collective behaviour at particular separation distances can give rise to resonant, very intense circularly polarized fields capable of uniformly enhancing CD spectroscopic signal of molecules over the entire extension of the array. Thus, in order to guarantee duality symmetry, we will operate at a particular frequency at which Kerker's first condition is fulfilled by every particle in the lattice, while we modify the separation distance between them to induce a resonant behaviour of the periodic system.

Under certain circumstances, the optical response of each nanoparticle can be described as a combination of a point electric dipole and a point magnetic dipole. Thus, in such situations, the collective behaviour of the system can be analyzed using an approach based on the Coupled Dipole Approximation (CDA). For a finite collection of $N$ particles, the optical response of the coupled electric and magnetic dipoles has a well-known algebraic solution:

\begin{equation}
\vec{\Psi} = \left[\bm{I} - k^2 \bm{G}_F \bm{\alpha}_F\right]^{-1} \vec{\Psi}_{0}.
\end{equation}
$\vec{\Psi}$ is a $6N$ dimensional vector which contains the self-consistent solution for each electric and magnetic field components in every site. $\bm{I}$ is the $6N \times 6N$ identity matrix and $\bm{\alpha}_F$ is a matrix of the same size containing the values of the electric and magnetic polarizabilities. $\bm{G}_F$ is a matrix that contains the coupling coefficients (built as lattice sums of Green's functions) between all the different dipoles in the system. Finally, $\vec{\Psi}_{0}$ is a $6N$ dimensional vector containing the values of the incident electric and magnetic fields at each of the positions of the particles (more detailed information can be found in the Supplementary Material).

Collective resonances emerge when $\text{det}\left( \bm{I} - k^2 \bm{G}_F \bm{\alpha}_F \right) \approx 0$ (notice that real poles would signal the existence of bound states). In these situations the scattered fields around the nanoparticles are strongly enhanced and, thus, $f_\text{CD}$ increases substantially. Note that since the incident wavelength and, therefore, the polarizabilities of the particles are fixed to fulfill duality symmetry, the emergence of this collective modes depends only on geometrical values, namely, the spatial distribution of the nanoparticles and the direction of incidence of the probing CPL.

It is important to note that there are different interaction mechanisms through which nanoparticles can interact in a periodic structure to give rise to collective resonances. On the one hand, if the photonic elements are placed at relatively long separation distances, they interact through a $r^{-1}$ dependent far-field dipolar interaction giving rise to diffractive modes ~\cite{hicks}. On the other hand, if the particles are more closely spaced, $r^{-3}$ and $r^{-2}$ dependent near-field interactions can also give rise to hybridized chain modes~\cite{ross,halas_review}.  

In what follows, we show that lattice resonances based on both types of interaction mechanisms are, in practice, useful to enhance the local density of optical chirality $C(\mathbf{r})$ and, thus, enhance the CD signal, setting a useful photonic paradigm for experimental testing. We have chosen our proof-of-concept system to be a finite 1D array of silicon dual nanoparticles showing that both types of lattice resonances give rise to an important enhancement of $f_\text{CD}$.

\section{\label{sec:level4}A finite chain of dual nanospheres}

In the following, we study a 1D array of $N=2000$ silicon nanospheres of radius $a = 150$ nm located along the $OY$ axis such that the nanospheres are placed at $\mathbf{r}_n = (n-1)D \mathbf{\hat{u}_y}$ where $n = 1,2,..., N$ and $D$ is the distance between two adjacent nanoparticles from center to center. In addition, the wavevector ($\mathbf{k}$) of the incoming circularly polarized plane wave is contained in the $YZ$ plane and forms an angle $\theta$ with the $OZ$ axis perpendicular to the chain. A sketch of the parameters of the system can be seen in Fig. \ref{Array}.

\begin{figure}[ht]
	\centering
	\includegraphics[width=1\textwidth]{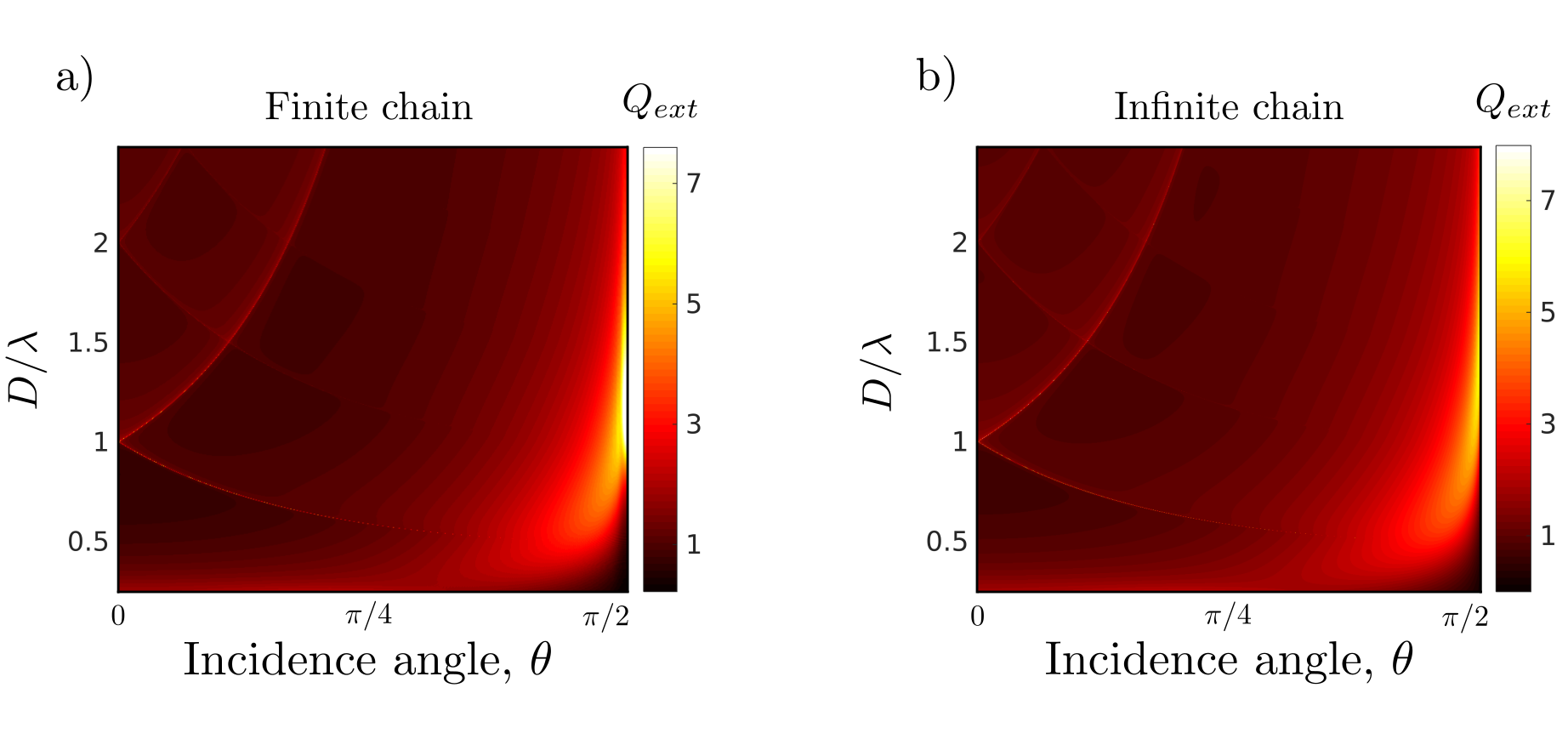}
	\caption{Normalized extinction cross section map under Kerker's first condition as a function of interparticle distance for: {\bf a)} a finite periodic 1D array of 2000 silicon nanospheres, {\bf b)} an infinitely periodic 1D array of silicon nanospheres.}
	\label{Diego_Extinction}
\end{figure}

To characterize the resonant features of the system, we first study the extinction cross section, $C_{ext}$ of the particle chain. In this particular system, Kerker's first condition is exclusively fulfilled at $\lambda = \lambda_k = 1208$ nm, so we explore the extinction cross section of the system as a function of two parameters: the incidence angle, $\theta$, and the interparticle distance, $D$. The results of this calculation are shown in Fig. \ref{Diego_Extinction}a, where the normalized extinction cross section, $Q_{ext} = C_{ext}/(N\cdot C_{ext,1})$, is shown and $C_{ext,1}$ is the extinction cross section of a single silicon particle at $\lambda_k$. In Fig. \ref{Diego_Extinction}b, a comparison of our results with the extinction cross section of an infinite chain of particles is shown (details of the theoretical formalism describing the infinite case can also be found in the Supplementary Material). As expected, both calculations give very similar results since collective interactions are known to converge for a given number of elements in a periodic structure~\cite{plasmon_pol, size_matters}. Interestingly, the number of nanospheres required in our system to reproduce the behaviour of the infinite case is very big comparing to what is found in the literature.

\begin{figure}[ht]
	\centering
	\includegraphics[width=1\textwidth]{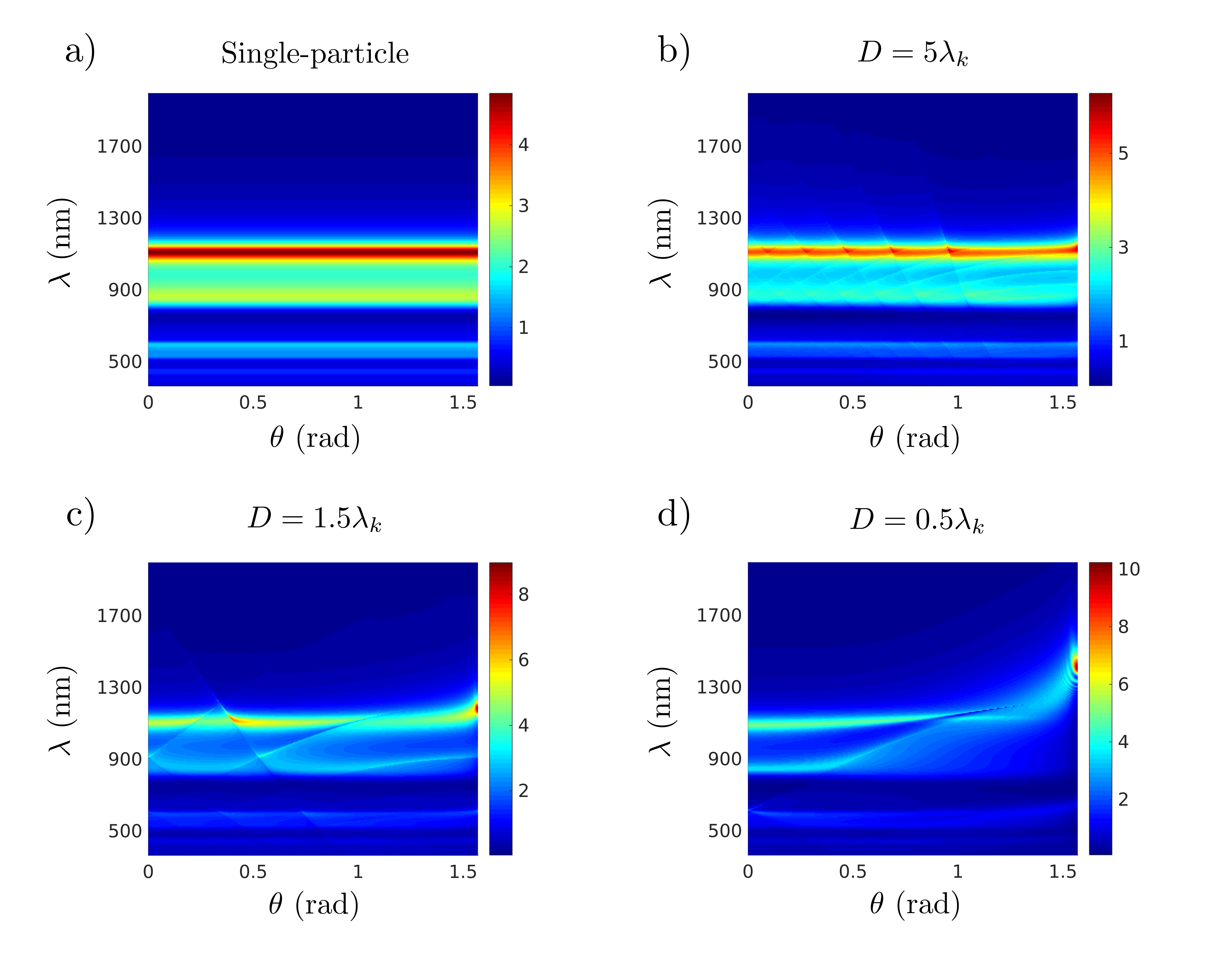}
	\caption{Normalized extinction cross section map as a function of the illuminating wavelength for: {\bf a)} a single silicon nanosphere, {\bf b)} a chain of 2000 nanospheres with $D = 5\lambda_k$, {\bf c)} a chain of 2000 nanospheres with $D = 1.5\lambda_k$ and {\bf d)} a chain of 2000 nanospheres with $D = 0.5\lambda_k$.}
	\label{LambdaSweep}
\end{figure}

The analytic calculations for the infinite chain allow us to identify the narrow peaks at low angles ($\theta \sim 0$) as diffractive modes emerging at $D/\lambda = m/(1 \pm \sin \theta)$ where $m$ takes positive integer values. As stated above, these peaks emerge as a consequence of the constructive interference of the far-field interactions. This is also observable in Fig. \ref{LambdaSweep}b, where, together with the single-particle spectrum, narrow diffraction lines emerge when nanoparticles are placed at $\lambda=5\lambda_k$.

However, close to grazing incidence ($\theta \rightarrow \pi/2$) a much wider resonance emerges. This mode does not fit with diffractive characteristics and, thus, has to be interpreted in terms of a different interaction mechanism. As we show in Fig. \ref{LambdaSweep}, this resonance emerges when particles in the chain are placed closer than $D = 1.5\lambda_k$ and, therefore, near- and mid-field interparticle contributions become important. As particles are placed closer together, the magnetic single-particle resonance of silicon ($\lambda_{mag} \sim 1110$ nm) redshifts and it can approach the wavelength of the Kerker condition, $\lambda_k$. All these considerations point out that this resonance is related to a \emph{bonding} transversal mode of the chain, a configuration which has been extensively studied in nanoparticle dimers and metamaterials~\cite{halas_review}.

Moreover, at $\theta = \pi/2$, the complete physical configuration (incident wave included) is rotationally symmetric. Due to the simultaneous duality and axial symmetry, helicity can be expressed in this situation as:

\begin{equation}
\Lambda = \frac{J_yP_y}{|P_y|},
\end{equation}
where both $J_y$ (y-component of the total angular momentum of light) and $\Lambda$ are conserved quantities. Therefore, $P_y$ (y-component of the linear momentum of light) cannot change sign under scattering and backward scattering is not allowed~\cite{kerker}. As a consequence, this mode in an exclusively forward propagating one.

Once we have thoroughly analyzed the optical resonances that can emerge in a dual 1D array of nanospheres, we proceed to show that, as predicted, these resonances give rise to important enhancements of $f_\text{CD}$. In order to model realistic environments of practical interest, we define two averaged enhancement factors: the average CD value around a particle in position $i$ ($f_{\text{CD}}^{i}$) and the array-averaged CD enhancement factor, $f_{\text{CD}}^{avg}$, defined as

\begin{eqnarray}
\label{fCDi}
f_{\text{CD}}^{i} &=& \int_{S_i}\frac{1}{4\pi} f_{\text{CD}}(\xi,\eta)\sin\xi d\xi d\eta, \\
\label{fCDavg}
f_{\text{CD}}^{avg} &=& \frac{1}{N}\sum_{i = 1}^N f_{\text{CD}}^{i}.
\end{eqnarray}
The integrals are carried out around a spherical surface $S_i$ ($\xi$ and $\eta$ are, respectively, the polar and azimuthal spherical coordinates respect to the center of the $i$-th nanoparticle) separated $1~\text{nm}$ apart from the surface of the $i$-th silicon nanoparticle.  

\begin{figure}[ht]
	\centering
	\includegraphics[width=0.92\textwidth]{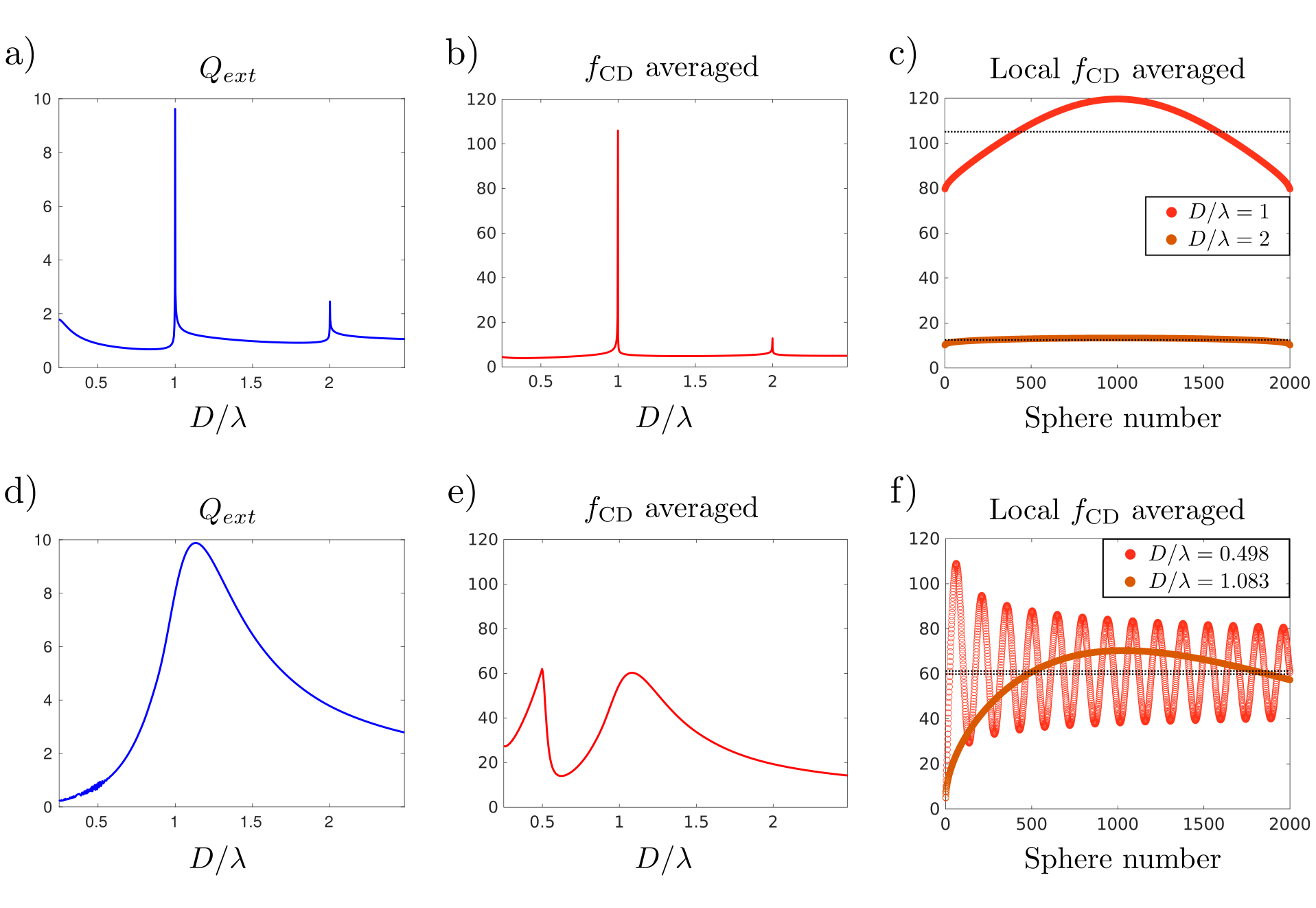}
	\caption{ {\bf a)} Normalized extinction for an incidence of $\theta = 0$. {\bf b)} $f_\text{CD}^{avg}$ for $\theta = 0$. {\bf c)} Distribution of the local CD enhancement factor for the two resonant conditions in b). {\bf d)} Normalized extinction for an incidence of $\theta = \pi/2$. {\bf e)} $f_\text{CD}^{avg}$ for $\theta = \pi/2$. {\bf f)} Distribution of the local CD enhancement factor for the two resonant conditions in e). Black dotted line: average enhancement for the whole structure ($f_\text{CD}^{avg}$) at the resonant conditions in Figs. {\bf b)} and {\bf c)}.}
	\label{Extinction_fCD}
\end{figure}

Fig. \ref{Extinction_fCD}a presents the normalized extinction cross section of the silicon array as a function of $D/\lambda$, for $\theta = 0$. Under this incidence, the extinction shows a resonant behaviour at the diffraction condition $D = m\lambda_k$, for $m \in \mathbb{N}$. However, the strength of the resonances decreases as $m$ is increased, something which is already noticeable in Fig. \ref{Extinction_fCD}a. In Fig. \ref{Extinction_fCD}b we can compare the extinction resonances with the CD enhancement averaged over the whole structure, $f_{\text{CD}}^{avg}$, for $\theta=0$. We obtain two orders of magnitude CD enhancement (a factor of $\sim 110$) averaged over the entire structure for the diffractive mode peaking at $D/\lambda=1$. Note that previous results for induvidual nanoparticles \cite{PRB} concluded that individual silicon nanoparticles of similar radius, were capable of inducing an enhancement of one order of magnitude in the molecular CD signal. Thus, this dual lattice design strategy provides an additional order of magnitude enhancement in $f_{\text{CD}}^{avg}$. 

Also, the CD enhancement variation among the individual spheres of the chain is shown in Fig. \ref{Extinction_fCD}c. In this figure, we reflect the $f_{\text{CD}}^i$ enhancement value around every sphere, for the two resonant distances appearing in Fig. \ref{Extinction_fCD}b. The average values of the resonant enhancement for the whole structure (coinciding with the peak values of Fig. \ref{Extinction_fCD}b) are presented as black dotted line. We can observe that both resonances have similar symmetric field distributions, whose maximum is located right at the center of the array.

On the other hand, under incidence parallel to the array axis, at $\theta = \pi/2$, the normalized extinction cross section in Fig. \ref{Extinction_fCD}d shows a single broad peak around $D/\lambda \sim 1.1$. Analyzing the array-averaged CD enhancement factor, though, two peaks are evident in Fig. \ref{Extinction_fCD}e, one at $D/\lambda \sim 1.1$ and another one around $D/\lambda \sim 0.5$, both showing maximum values of $f_\text{CD}^{avg} \sim 60$. As pointed out above, the $f_\text{CD}^{avg}$ peak centered at $D/\lambda \sim 1.1$ is related to the redshifted single-particle magnetic resonance and, thus, it is directly related to the the peak in Fig. \ref{Extinction_fCD}d.

However, the second $f_\text{CD}^{avg}$ peak at $D/\lambda \sim 0.5$ is not related to an extinction peak, but to the small oscillations that the normalized extinction shows in Fig. \ref{Extinction_fCD}d for $D/\lambda \le 0.5$. This oscillations do not appear in the infinite case, which allows us to understand it as a finite chain effect. Finite chain modes have been widely studied over the literature and are known to appear for the geometric condition $D/\lambda < 0.5$ (concretely for $D/\lambda = \frac{(N-2)n + 1}{2N(N-1)}$, with $n \in [1,2,..., N]$)~\cite{finite_chain}, which very well fits the range in which both the extinction oscillations and the $f_\text{CD}^{avg}$ peak are found. The nature of these modes makes them have nearly $0$ dipole moment, which implies that they can barely contribute to the extinction but can still present very strong near-fields~\cite{DarkPlasmon}, giving rise to the second peak in Fig. \ref{Extinction_fCD}e. This happens because finite chain modes are typically characterized by sign changing dipole distributions~\cite{finite_chain}, whose contributions interfere destructively in the far-field. Figure \ref{Extinction_fCD}f shows the spatial distribution of $f_\text{CD}^{avg}$ in the chain for the two resonant conditions in Fig. \ref{Extinction_fCD}e. Finally, note that the $f_\text{CD}$ distribution for the finite chain mode in Fig. \ref{Extinction_fCD}f represents a guided mode of the finite chain, with a considerable propagation length.

In conclusion, we have presented a systematic procedure to design periodic nanophotonic platforms capable of enhancing molecular CD spectroscopy resonantly. As an example, we applied the method to design a dual 1D periodic nanostructure made of high refractive index nanoparticles. By analyzing the emergent lattice resonances in the system we have found that they can be classified in three different types: diffractive modes, non-diffractive modes and finite chain modes. We have shown that far-field diffractive resonances can increase the array-averaged CD enhancement up to a factor $\sim 110$, providing an additional order of magnitude enhancement compared to the optical response of a single particle. Moreover, we find that non-diffractive and finite chain modes, which have not been previously considered in the field-enhanced CD literature, also give rise to comparable enhancement magnitudes (over a factor of $\sim 60$). This method can be easily extended to more complicated platforms such as 2D metasurfaces or 3D photonic crystals, opening venues to new designs and research directions.

\bibliography{main}

\end{document}